\documentclass[reprint,amsmath,amssymb,aps]{revtex4-2}
\usepackage{graphicx}
\usepackage{dcolumn}
\usepackage{bm}
\usepackage{physics}
\usepackage{hyperref}
\usepackage[mathlines]{lineno}

\begin{document}

\preprint{APS/123-QED}

\title{High-Fidelity Transport of Trapped-Ion Qubits in a Multi-Layer Array}

\author{Deviprasath Palani, Florian Hasse, Philip Kiefer, Frederick Boeckling, Jan-Philipp Schroeder, Ulrich Warring, Tobias Schaetz}

\affiliation{Albert-Ludwigs-Universitaet Freiburg, Physikalisches Institut Hermann-Herder-Strasse 3, Freiburg 79104, Germany}

\date{\today}

\begin{abstract}A variety of physical platforms are investigated for quantum control of many particles, and techniques are extended to access multiple dimensions.
Here, we present our experimental study of shuttling single Mg$^+$ ions within a scalable trap-array architecture that contains up to thirteen trapping sites in a three-dimensional arrangement.
We shuttle ions from a dedicated loading hub to multiple sites with a success rate of larger than $0.99999$. 
In a prototype application, we demonstrate the preservation of the coherence of superposition states of a hyperfine qubit during inter-site shuttling.
Our findings highlight the potential of these techniques for use in future large-scale architectures.
\end{abstract}

\maketitle

%
%
Developments toward larger-scale quantum technology are driven by conceived performance advantages for quantum information processing, experimental simulation, and metrology applications\,\cite{Ladd2010, georgescu_quantum_2014, Degen2017}.
Experimental research pursues two common objectives in multiple eligible platforms: (i) improving manipulation techniques towards near-unity fidelity, and (ii) fostering architectures with global and local control capabilities.
Trapped atomic ions naturally feature reproducible qubits and demonstrate notable performance qualities, approaching operational fidelities for all required tasks at a level of 0.9999 for ensembles of few-ions in linear arrangements\,\cite{Erickson2022, Srinivas2021, Brewer2019, Monroe2021, Ballance2016, Harty2014}.
Integration of these features into sufficiently extendable architectures is investigated in several ways, e.g., following bottom-up approaches of quantum charged-coupled devices (QCCD)\,\cite{wineland_experimental_1997, Kielpinski2002}, ion-photon networks\,\cite{Monroe_ionphoton_2014, Inlek_multispecies_2017}, and trapped-ion arrays\,\cite{schmied_optimal_2009, sterling_fabrication_2014, mielenz_arrays_2016, bruzewicz_scalable_2016} or top-down approaches\,\cite{Gilmore2021, Marissa2021, Joshi2020}.
The QCCD, on the one hand, builds on ion qubits stored in memory regions, being shuttled along dedicated ion-guides to interaction regions for logic operations, e.g., to build up many-qubit entanglement\,\cite{Pino2021}.
Required shuttling routines are performed, mitigating ion-loss\,\cite{Blakestad_transport_2009} while being optimized for speed\,\cite{Bowler_transport_2012,walther_transport_2012, Sterk_transport_2022} within interlinked 1D trapping arrays.
On the other hand, the array of individually trapped ions features trapping sites at sufficiently small distances that allow for directly exploiting the Coulomb interaction at long range.
This may allow evolving the quantum system as a whole, even in higher dimensions.
For example, multidimensional entanglement is envisioned to be conceived by engineered and tunable inter-site interactions -- required methods are established in one\,\cite{brown_coupled_2011, harlander_trapped-ion_2011, wilson_tunable_2014} and two dimensions\,\cite{mielenz_arrays_2016, Hakelberg2019, Kiefer2019, Warring2020}.
The latter is anticipated to permit addressing currently intractable open questions of interest via experimental quantum simulations\,\cite{cirac_goals_2012, Warring2020}.
Combining the advantages of QCCD and multidimensional array architectures may turn out beneficial for future quantum applications.

In this Letter, we present high-fidelity shuttling of single Mg$^+$ between different sites of our multi-layer (2d+) trap-array.
The success rate for shuttling through locally controlled transfer channels is larger than $0.99999$.
We observe an excess motional mode heating smaller than 150\,quanta for a single shuttle operation.
In addition, we preserve coherence of the electronic degree of freedom despite shuttling ions through four sites of an array spanning a triangular-pyramidal configuration.
%
%
\begin{figure}[t]
\includegraphics{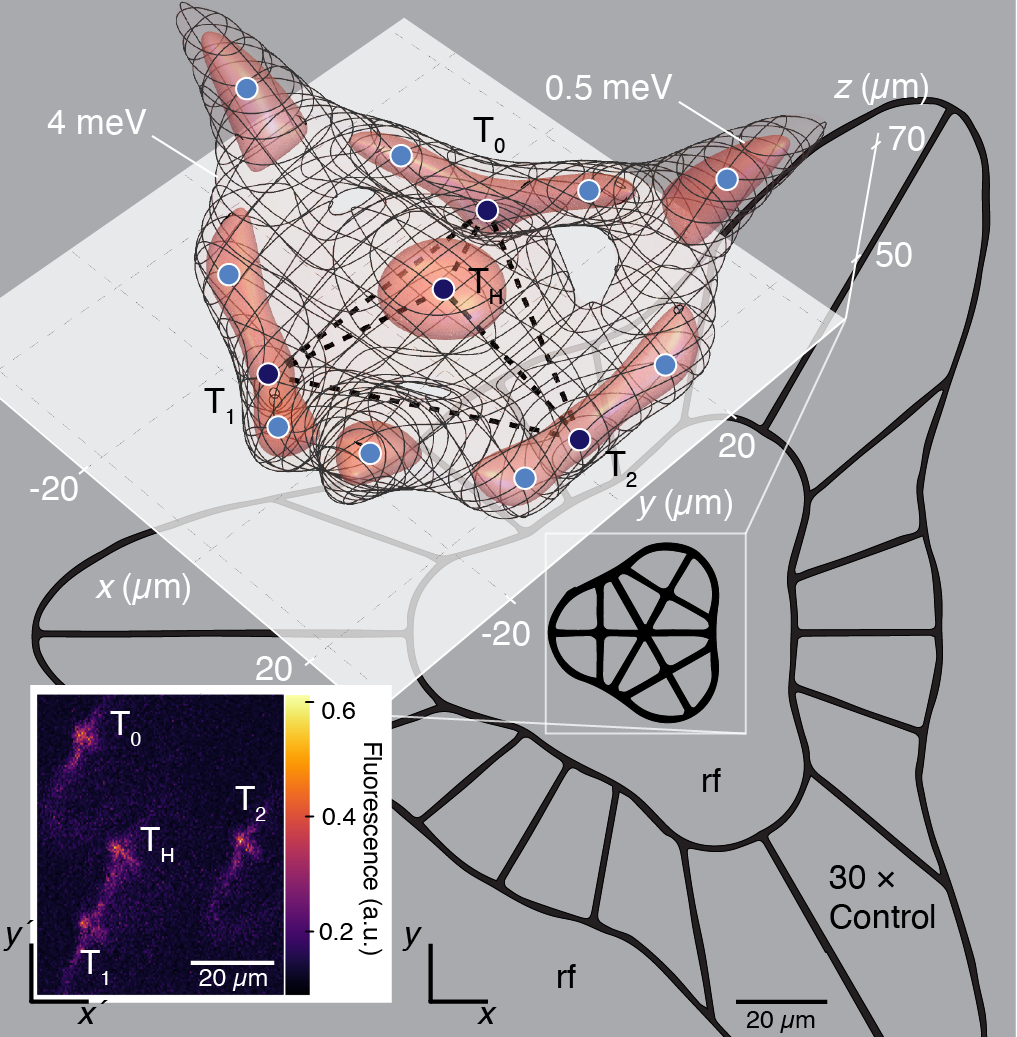}%
\caption{\label{fig1}Illustration of our trap-array archeticture.
In the background, we show a schematic of the two radio-frequency and 30 control electrodes used for the generation and fine control of our trapping landscape.
Overall, we create 13 discrete trapping sites marked by disks in the overlay.
We depict calculated pseudo-potential equipotential surfaces for $0.5$\,meV (red, high opacity), mutually separating seven regions with single or multiple sites, and $4$\,meV (red, low opacity). 
To effectively open and close inter-regional ion transport channels, we switch between dedicated control potentials.
We highlight four sites that define a unit cell of a 2d+ lattice in the shape of a triangular pyramid (dashed lines).
The inset shows a false-color fluorescence image of four single ions at T$_{\text{0}}$, T$_{\text{1}}$, T$_{\text{2}}$, and T$_{\text{H}}$, respectively. 
Ions are observed at a $16^\circ$ angle relative to the surface normal, and their positions are adjusted by local control fields.
}
\end{figure}
%

%
%
%
%
Our custom-designed ultra-high vacuum chamber hosts the radio-frequency (rf) surface-electrode trap, featuring an array of individual trapping sites, fine-tuned by 30 discrete control electrodes\,\cite{mielenz_arrays_2016}, see Fig.\,\ref{fig1}. 
We employ laser beams aligned parallel to the chip surface from five different laser sources for ablation loading\,\cite{Leibrandt_ablation_2007, Zimmermann_ablation_2012} and manipulation\,\cite{leibfried_quantum_2003} of $^{24}$Mg$^+$ and $^{25}$Mg$^+$ (with hyperfine structure) in a quantization field $|\mathbf{B}_\text{quant}|\simeq 10.9$\,mT\,\cite{hakelberg_hybrid_2018}.
We find average storage duration (1/e time) of single ions to be $\simeq15$\,minutes. 
Currently, storage is limited by background gas collisions, consistent with the residual gas pressure of our chamber at $\simeq10^{-9}$\,Pa.
We image the fluorescence light of trapped ions either on an electron-multiplying charge-coupled device camera or a photo-multiplier tube (PMT).
Our experimental control hardware is software interfaced by Python, and we control parameters, record data, and perform data analysis routines in a JupyterLab environment. 

%
%
The trap chip is fabricated by Sandia National Laboratories\,\cite{mielenz_arrays_2016}.
A high-power frequency generator provides a power of $\simeq26$\,dBm oscillating at $\Omega_{\text{RF}}/(2\pi)\simeq52.4\,\text{MHz}$ to a helical resonator that amplifies the trapping voltage with a loaded quality factor of $\simeq20$.
The three-dimensional trapping landscape of the array is generated by a common rf potential applied to two optimally shaped electrodes\,\cite{mielenz_arrays_2016}. We estimate a rf voltage amplitude of $U_{\text{RF}}\simeq200$\,V (zero-to-peak).
In Fig.\,\ref{fig1}, we show two calculated equipotential surfaces of the corresponding pseudo-potential $\Phi_{\text{rf}}$ for $^{24}$Mg$^+$ at 0.5\,meV and 4\,meV, respectively.
In this pure rf configuration (all other electrodes are electrically grounded), we create seven distinct trapping regions featuring 13 sites, each site is highlighted by disks in the overlay of Fig\,\ref{fig1}. 
The regions are mutually separated by the 0.5\,meV potential barrier, but overall enclosed by the 4\,meV potential barrier. 
In the following, we focus on three in-plane (the $xy$ plane, parallel to the surface) sites and one out-of-plane site: 
The three sites, labelled T$_{\text{0}}$, T$_{\text{1}}$ and T$_{\text{2}}$, form a $40\,\mu$m equilateral triangle at $z\simeq40\,\mu$m. 
The site T$_{\text{H}}$ (hub) is situated at the center of our array, but is elevated to $z\simeq53\,\mu$m. 
In combination, all four sites define a unit cell in the shape of a triangular pyramid.
For our experimental parameters, we find mode frequencies $\omega/(2\pi)$ of 4 to 7\,MHz of single ions in sites T$_{\text{0}}$, T$_{\text{1}}$, and T$_{\text{2}}$ and of 1 to 3\,MHz in T$_{\text{H}}$.

A metal sheet mounted $\simeq7$\,mm above the chip surface is biased to apply a global electric field to compensate stray fields along the $z$-axis.
In addition, we use 30 control electrodes (integrated in the chip) for local fine-tuning of trapping conditions.
For example, we apply local stray field compensation and tuning of local potential curvatures, i.e., mode frequencies and orientations.
Control voltages are supplied by a 36-channel arbitrary waveform generator (AWG)\,\cite{bowler_arbitrary_2013} with outputs connected via low-pass filters with cut-off frequencies $\omega_\text{lowpass}/(2\pi)\simeq7$\,kHz. 
The AWG is integrated into our data acquisition system for real-time playback of waveforms within our experimental sequences with a timing accuracy of $10$\,ns.
In this work, we use six dedicated control potential configurations $\Phi_{\text{c}}$, designated and calibrated for ion loading $\Phi_{\text{HL}}$ and storage $\Phi_{\text{H}}$ in T$_{\text{H}}$, as well as, single-ion storage in one of the three lower lying sites: $\Phi_{\text{0}}$, $\Phi_{\text{1}}$, and $\Phi_{\text{2}}$.
Moreover, we calibrated $\Phi_{\text{Pyramid}}$, to initialize the pyramid, i.e., single ions in all four sites, respectively.
Note that the total trapping potential is given by $\Phi_{\text{rf}}+\Phi_{\text{c}}$ (here, ignoring stray potential contribution), and we describe our routine to find such configurations in\,\footnote{See Supplemental Material}. 

To shuttle single ions, we switch between initial and final $\Phi_{\text{c}}$ with variable timings $t_{\text{playback}}$ of 0.1 to 1\,ms.
For example, to shuttle the ion from T$_{\text{H}}$ to T$_{\text{1}}$, we choose initially $\Phi_{\text{c}} = \Phi_{\text{H}}$ and switch to $\Phi_{\text{c}} = \Phi_{\text{1}}$.
During playback of the corresponding waveform comprised by a linear ramp, a transfer channel between T$_{\text{H}}$ and T$_{\text{1}}$ is established, and the ion is shuttled.
To reverse this process, we shift back to $\Phi_{\text{c}} = \Phi_{\text{H}}$.
In this way, we can engineer arbitrary inter- and intra-regional shuttling, while we focus here on demonstrations of inter-regional shuttling.

%
%
For loading ions, we use an ablation laser with a wavelength of $1030$\,nm, a repetition rate of $1.5$\,kHz, a pulse width of $1.4$\,ns, and a maximal pulse energy of $93\,\mu$J.
The beam is focused by a $150$\,mm focal length lens onto a Mg wire with 0.4(1) mm diameter inside the vacuum chamber, mounted at a distance of $\simeq10$\,mm from T$_{\text{H}}$.
We tune the pulse energy by a variable neutral density filter and estimate the beam waist on the target to be smaller than 1/10 of the diameter of the wire.
A photo-ionization beam near $285$\,nm is tuned to selectively ionize $^{24}$Mg or $^{25}$Mg. 
We pass the beam near T$_{\text{H}}$ with a waist of $30\,\mu$m and a power of $\simeq1$\,mW.

Three additional laser beams from a common laser source are tuned to address a S$_{1/2}$ to P$_{3/2}$ cycling transition with natural line width $\Gamma/(2\pi)\simeq42$\,MHz: Two for Doppler cooling\,\cite{leibfried_quantum_2003}, detuned by $-8\,\Gamma$ and $-\Gamma/2$, and one tuned close to resonance for detection.
Optionally, we use two additional laser beams from another source tuned to address dedicated S$_{1/2}$-P$_{1/2}$ transitions for improved state preparation in $^{25}$Mg$^+$ via optical pumping.
Laser frequencies of both sources are stabilized via a wavemeter and/or Doppler-free saturation absorption spectroscopy.
All five laser beams (for global ion manipulation) are superimposed, aligned with $\mathbf{B}_\text{quant}$, and polarized to yield $\sigma^+$ polarization. 
We align these beams with waists $\simeq30\,\mu$m at T$_{\text{H}}$. 
This results in significantly lower beam intensities at sites T$_{\text{0}}$, T$_{\text{1}}$, and T$_{\text{2}}$, which we estimate at 25\,$\%$, 30\,$\%$, and 15\,$\%$, respectively. 
Optionally, we apply local AC Stark shifts\,\cite{leibfried_quantum_2003} via a dedicated laser beam detuned by $\simeq+1000\,\Gamma$ from the S$_{1/2}$-P$_{3/2}$ transition. 
This far-detuned beam is focused at T$_{\text{1}}$ with a waist of $\simeq10\,\mu$m. 

Using $^{25}$Mg$^+$, we span a qubit by the hyperfine ground states $\ket{\text{S}_{1/2}, \text{F}=3, \text{m}_\text{F}=+1}\equiv \ket{\downarrow}$ and $\ket{\text{S}_{1/2}, \text{F}=2, \text{m}_\text{F}=0}\equiv	 \ket{\uparrow}$ coupled via a first-order field-independent transition $\omega_\text{qubit}/(2\pi)\simeq1762.974$\,MHz at $|\mathbf{B}_\text{quant}|$\,\cite{hakelberg_hybrid_2018}.
We can coherently manipulate all S$ _{1/2}$ hyperfine states via microwaves. 
We currently prepare qubit states with a fidelity of 90(2)$\%$ and find a coherence time $\tau_{\text{coh}}\simeq6$\,ms for equal superposition states of $\ket{\downarrow}$ and $\ket{\uparrow}$. 
State preparation fidelity and coherence duration are likely to be limited by stray currents in rf electrodes.
Further, we measure static magnetic-field gradients of $\simeq0.013\,\text{T}/\text{m}$ near T$_{\text{H}}$.

An objective, collecting fluorescence light from the ions, is mounted at an angle of $\simeq16^\circ$ regarding the $z$-axis and projects tilted image planes onto our camera chip. 
We can spatially resolve individually trapped ions with a resolution of $0.32(1)\,\mu$m per pixel.
The inset of Fig.\,\ref{fig1} shows a false-color image of recorded fluorescence light of four $^{24}$Mg$^+$ forming a fully filled unit cell. 
Ion positions are altered from the ideal triangular-pyramidal locations by less than $5\,\mu$m -- considering a correction for the tilted perspective.
Alternatively, we collect photon count histograms with the PMT, and estimate motional mode excitation\,\cite{kalis_motional-mode_2016} or electronic-state populations\,\cite{leibfried_quantum_2003}.

\begin{figure}[]
\includegraphics{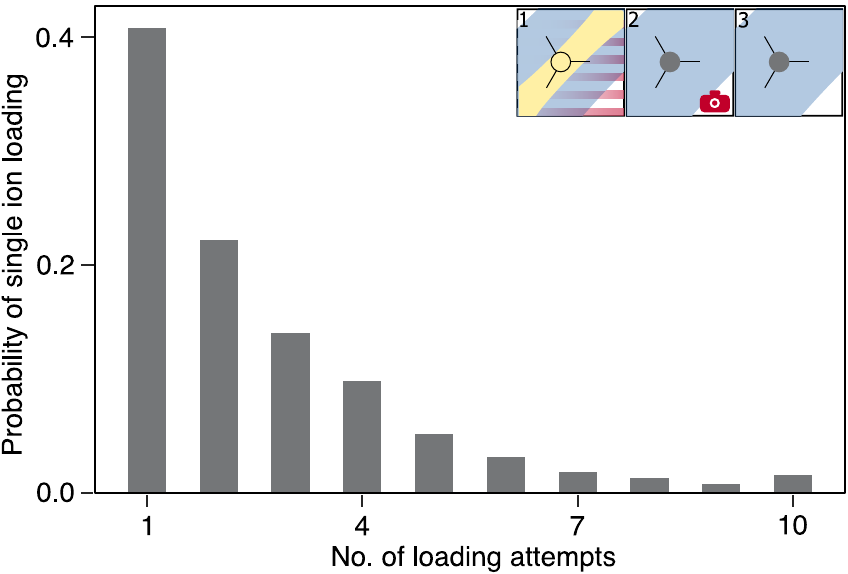}
\caption{\label{fig2} Automated loading of single ions at the hub for subsequent redistribution within the unit cell.
(Inset) A single loading attempt is comprised by the following sequence:
1) Ablation (red, five to ten pulses), photo-ionization (yellow), and cooling lasers (blue), as well as control potentials, are engaged for loading in T$_{\text{H}}$, 2) cooling laser remains on during a triggered snapshot of our camera, the image is analyzed for the presence of a single ion, and 3) a successfully loaded ion is kept Doppler cooled.
Typically, we use about five to ten laser ablation pulses in a single loading attempt.
The histogram shows the accumulated 973 attempts of $\simeq180$\,days of data taking. 
}
\end{figure}

%
%
In most of our experiments presented here, we utilize an automated single-ion loading sequence. 
We study its repeatability with data taken over the course of about 180 days (see Fig.\,\ref{fig2}).
%
%
We choose a pulse energy of the ablation laser of $\simeq30\,\mu$J.
The automated sequence, illustrated in the inset of Fig.\,\ref{fig2}, comprises three steps. 
Step\,1: Control voltages are set to $\Phi_{\text{HL}}$, cooling and photo-ionization beams are on, while a burst of about five to ten ablation pulses is fired.
Step\,2: The program captures a single camera frame with 500\, ms integration duration, and analyzes the image to identify the presence of a single ion at T$_{\text{H}}$.
Steps\,1 and 2 take in total about $1.2$\,s. 
If a single ion is detected, step 3 is engaged: The AWG is triggered to switch to $\Phi_{\text{H}}$ and the ion is continuously Doppler cooled.
If the loading attempt is unsuccessful, the program repeats steps\,1 and 2. 
In case ten subsequent loading attempts are unsuccessful, the automated loading sequence is terminated, and we manually adjust relevant parameters, e.g., re-tuning of $\Phi_{\text{HL}}$.
During about 180\,days, we attempt 973 loading bursts, and we find a probability of about $\simeq0.4$ to be successful after the first burst.
Typically, we carefully check relevant experimental parameters once per week.
For loading of multiple ions, we adapt the sequence to fill dedicated sites. For example, for filling all four sites of the pyramid (see inset of Fig.\,\ref{fig1}), we repeat steps\,1 and 2 until we load four ions into T$_{\text{H}}$ and switch control potentials from $\Phi_{\text{HL}}$ to $\Phi_{\text{Pyramid}}$. 

\begin{figure}
\includegraphics{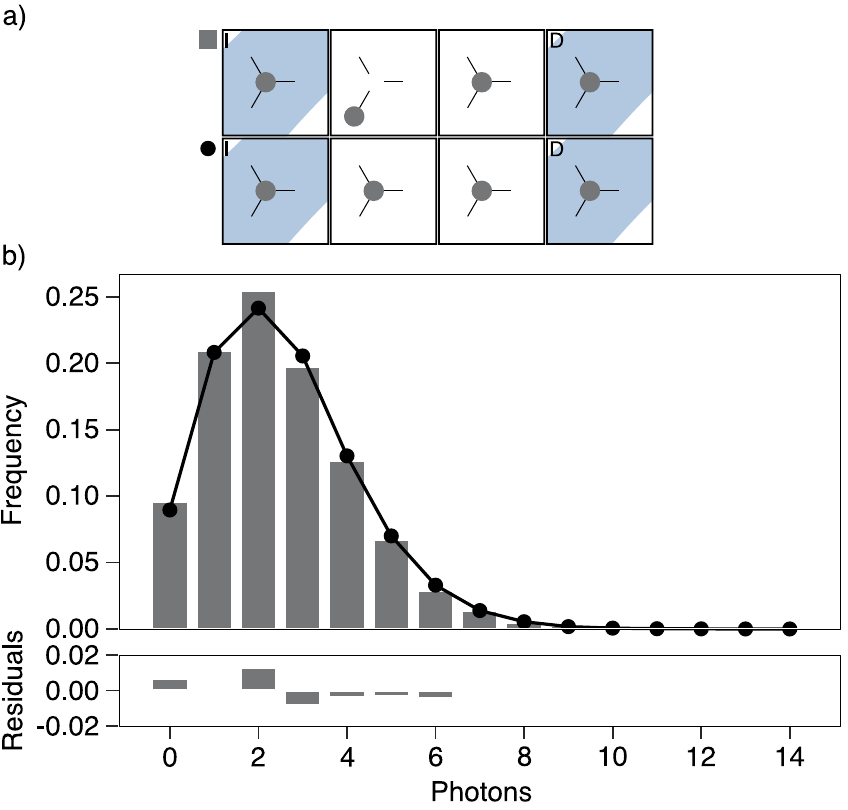}
\caption{\label{fig3}
Quality of shuttling single ions from T$_{\text{H}}$ to T$_{\text{1}}$ and back (roundtrip). 
We enable transport by playback of waveforms between fine-tuned control potential configurations $\Phi_\text{H}$ and $\Phi_\text{1}$ within $0.1\,$ms. 
During reference experiments, we fix the control potential configuration to $\Phi_\text{H}$. 
In a), we illustrate transport (squares) and reference (discs) sequences that I) initialize and D) detect $^{24}$Mg$^+$ at T$_{\text{H}}$. 
Here, we execute 11,000 repetitions of shuttling without ion loss.
In b), we show corresponding photon count histograms (reference: disks joined by black line) and find a fluorescence change of $-2.1(6)\,\%$ due to shuttling.
We summarize relocation results to all sites in Table\,\ref{tab:table1}, and find that mode heating due to shuttling is $<300$\,quanta per round trip.
}
\end{figure}

To benchmark the shuttling fidelity of single ions, we study relocation roundtrips with $^{24}$Mg$^+$.
The experimental sequence (Fig.\,\ref{fig3}a) comprises initialization and detection at T$_\text{H}$ with interleaved shuttling roundtrips to T$_\text{1}$ at a distance of $d_\text{H,1}\simeq25\,\mu$m. 
We measure the change of average fluorescence rate due to a shuttling roundtrip in comparison to reference histograms taken for the ion remaining at T$_\text{H}$.
In Fig.\,\ref{fig3}b, we show corresponding data for 11,000 repetitions.
In this way, we calculate an upper bound for the motional heating of 84 quanta per round trip, estimated for the lowest-frequency mode at $2\pi\,1\,$MHz, to get a conservative bound.
Note, that excitations are more-likely to be unevenly distributed on all three modes, and details depend on the evolution of mode orientations regarding rf field lines along shuttling trajectories.
While we assume in the following that excitations are non-coherent (thermal) for our shuttling speeds $d_\text{H,1}/t_\text{playback}\simeq0.25$\,m/s, we attribute excess heating to a rf noise heating mechanism and digital sampling noise\,\cite{Blakestad_transport_2009, Blakestad_transport_2011}.
%
We summarize corresponding shuttling results to all sites in Tab.\,\ref{tab:table1}, and emphasize that motional-heating rates per shuttling roundtrip are $<300$\,quanta.
The average ion storage duration, currently, limits us to $\simeq100,000$ consecutive shuttling sequences, yielding an estimate for a shuttling failure rate of lower than $10^{-5}$.
\begin{table}
\caption{\label{tab:table1} 
Quality of roundtrips from T$_{\text{H}}$ to all sites, following similar protocols as illustrated in Fig.\,\ref{fig3}a.
We list fluorescence changes, upper bound estimates of mode heating per roundtrip, and maximal slew rates required for the playback of waveforms.}
\begin{ruledtabular}
\begin{tabular}{cccc}
Visited &$\Delta$Fluo. per  &Mode heating &Slew rate\\
site & round trip($\%$) &(quanta/roundtrip) &(V/ms) \\
\hline
$T_{0}$ &-8.7(6) &$\lessapprox$287 &$<$4\\
$T_{1}$ &-2.1(6) &$\lessapprox$84 &$<$3\\
$T_{2}$ &-5.9(6) &$\lessapprox$197 &$<$3\\
\end{tabular}
\end{ruledtabular}
\end{table}

\begin{figure}
\includegraphics{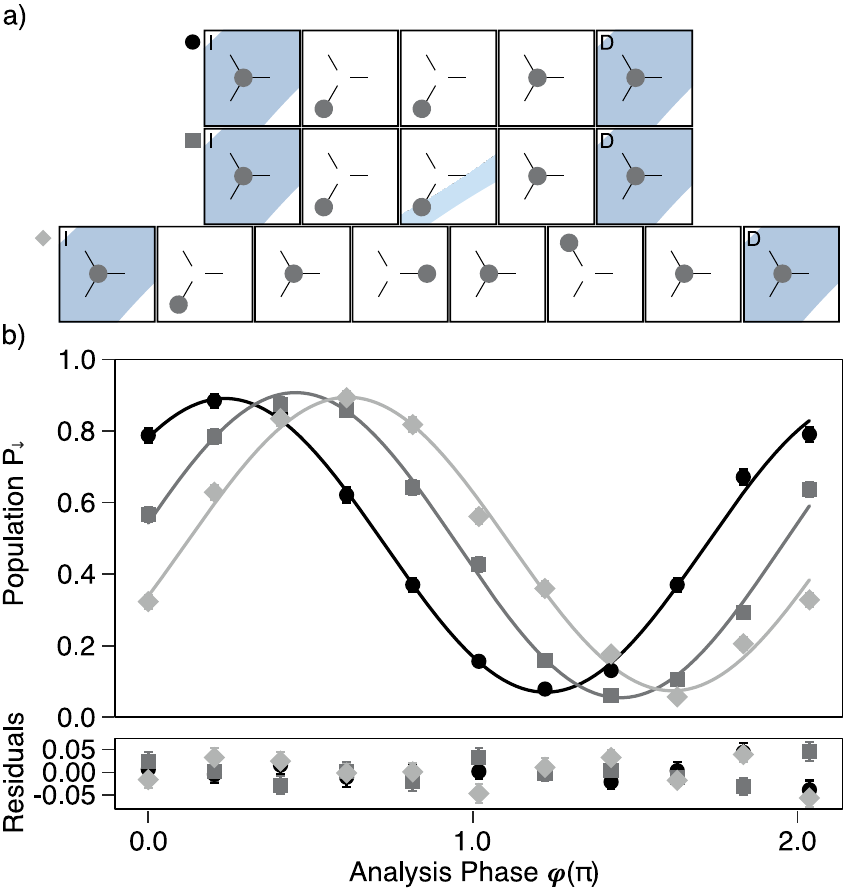}
\caption{\label{fig4}Coherent manipulation of qubit states in $^{25}$Mg$^+$ with multiple relocations. 
a) In three single-ion sequences, we initialize I) $\ket{\downarrow}$-$\ket{\uparrow}$ superposition and probe coherences with analysis pulses before D) detection in T$_{\text{H}}$. 
In the first and second sequences, we probe single relocation roundtrips to T$_\text{1}$ (discs) and apply optional local AC Stark shifts (squares). 
In the third sequence, we study the effect of multiple subsequent relocation roundtrips to all sites (diamonds).
In b), we compare the resulting P$_\downarrow$ for all sequences as a function of analysis phase $\varphi$.
Model fits result anticipated phase shifts and probed coherence amplitudes are consistent with background effects.
}
\end{figure}

To study the effects of shuttling on the electronic degrees of freedom, we probe the coherence of qubit superposition states in $^{25}$Mg$^+$ for multiple ion relocation scenarios. 
All sequences are comprised of three steps, see Fig.\ref{fig4}a.
Step\,1: We Doppler cool, and initialize an equal superposition of $\ket{\downarrow}$ and $\ket{\uparrow}$ with a microwave $\pi$/2 pulse at T$_\text{H}$.
Step\,2: We implement this step in three different versions for direct comparison: 
i) Ion is shuttled to T$_\text{1}$ and back, 
ii) ion is shuttled to T$_\text{1}$, and we apply a calibrated AC stark shift with the far-detuned beam for $50\,\mu$s before shuttling it back, 
iii) we engage multiple subsequent shuttling roundtrips from T$_\text{H}$ to all sites. 
Note, versions i) and ii) span a total duration of $0.25$\,ms and iii) spans $0.6\,$ms, i.e, 24 to 10 times shorter than $\tau_{\text{coh}}$.
Step\,3: Before fluorescence detection at T$_\text{H}$, we use an analysis $\pi$/2 pulse of variable phase $\varphi$, relative to the initial $\pi$/2 pulse, to probe the qubit coherence. 
We analyze photon count histograms to estimate the population probability P$_\downarrow$ of state $\ket{\downarrow}$, and plot data points as a function of $\varphi$ in Fig.\,\ref{fig4}b.
Each data point is an average of 1,000 repetitions.
Using sinusoidal model fits, we extract the contrast and phase offsets.
We compare our findings to the results of dedicated reference sequences, in which a single ion remains at T$_\text{H}$:
The contrasts are $82(1)\,\%$, $85(2)\,\%$, and $82(3)\,\%$, respectively, while for the reference on average $83(1)\,\%$ -- coherence rates are consistent with the reference.
Further, we find phase offsets relative to corresponding reference sequences of $5(4)^\circ$, $44(3)^\circ$, and $4(4)^\circ$, respectively. 
Relative shifts arise from the magnetic-field gradients and are, in particular, in agreement with the anticipate AC-stark shift of 2$\pi\,$2.21(6)\,kHz applied during scenario ii). 
We conclude that dephasing contributions due to ion shuttling are below our resolution, while relative phase shifts can be accounted for via appropriate bookkeeping.
%
%
%

%
%
To summarize, in approximately $90\,\%$ of the attempts, we load and confirm the presence of single ions within 6\,seconds at any site on demand. 
Simultaneous deterministic loading of four sites with single ions is enabled by mutual Coulomb repulsion. 
Alternative loading methods, e.g., using a pre-cooled atom source\,\cite{MOT_loading_2007, MOT_loading_Chiaverini_2012} may be beneficial in refilling arrays even more efficiently.
After loading, a single ion is shuttled to four sites and back for more than 100,000 repetitions (limited by background gas pressure) with a mean energy gain of below 150 quanta per trip within a single trip duration of $0.1\,$ms.
Tuning of experimental parameters can enable even non-adiabatic shuttling entailing moderate motional heating\,\cite{Bowler_transport_2012, walther_transport_2012, Blakestad_transport_2011}.
Previously, we showed the preparation of quantum mechanical ground states in single and multiple sites\,\cite{kalis_motional-mode_2016}, control of motional parameters\,\cite{mielenz_arrays_2016}, and tunable inter-site interactions via engineered phonon couplings\,\cite{Hakelberg2019, Kiefer2019} in two dimensions.
On the one hand, establishing phonon couplings between arbitrary sites in our 2d+ array permits accessing quantum simulations extended in size and dimension\,\cite{schaetz_towards_2007}.
On the other hand, exploiting two-to-three times smaller distances $d$ (interaction strength $\propto 1/\text{d}^3$) and ancilla ions (boosting interactions by additional electric charges), coupling rates can significantly exceed current decoherence rates. 
Technical limitations such as electric-field noise and background gas pressure can be mitigated by surface treatment methods\,\cite{hite_100-fold_2012} and cryogenic environments\,\cite{labaziewicz_suppression_2008}.
Finally, we substantially extend, here, our toolbox by shuttling qubits within our array, while preserving coherence.
Thus, we demonstrate prototype features of a 2d+ QCCD that can be upgraded with existing tools\,\cite{Kielpinski_sympathetic_2000,chen_sympathetic_2017,barrett_sympathetic_2003, Pino2021}.
Our architecture, with its tunable long-range interactions, holds potential for future exploitation. 
By combining it with the exceptional operational fidelities of single and few ions systems, we may be able to initialize and further explore multi-dimensional, fiducial entangled states.

%
\begin{acknowledgments}
The trap chip was designed together with R. Schmied in cooperation with the NIST ion storage group and produced by Sandia National Laboratories. 
This work was supported by the Deutsche Forschungsgemeinschaft (DFG) (Grant No. SCHA 973/6-3) and the Georg H. Endress foundation.
We thank Joern Denter, Kambiz Mahboubi, Fraenk Grossman, and their teams for technical support.
\end{acknowledgments}
%

%
\newpage
\section*{Supplemental Material: High-Fidelity Transport of Trapped-Ion Qubits in a Multi-Layer Array}

\begin{figure}
\includegraphics{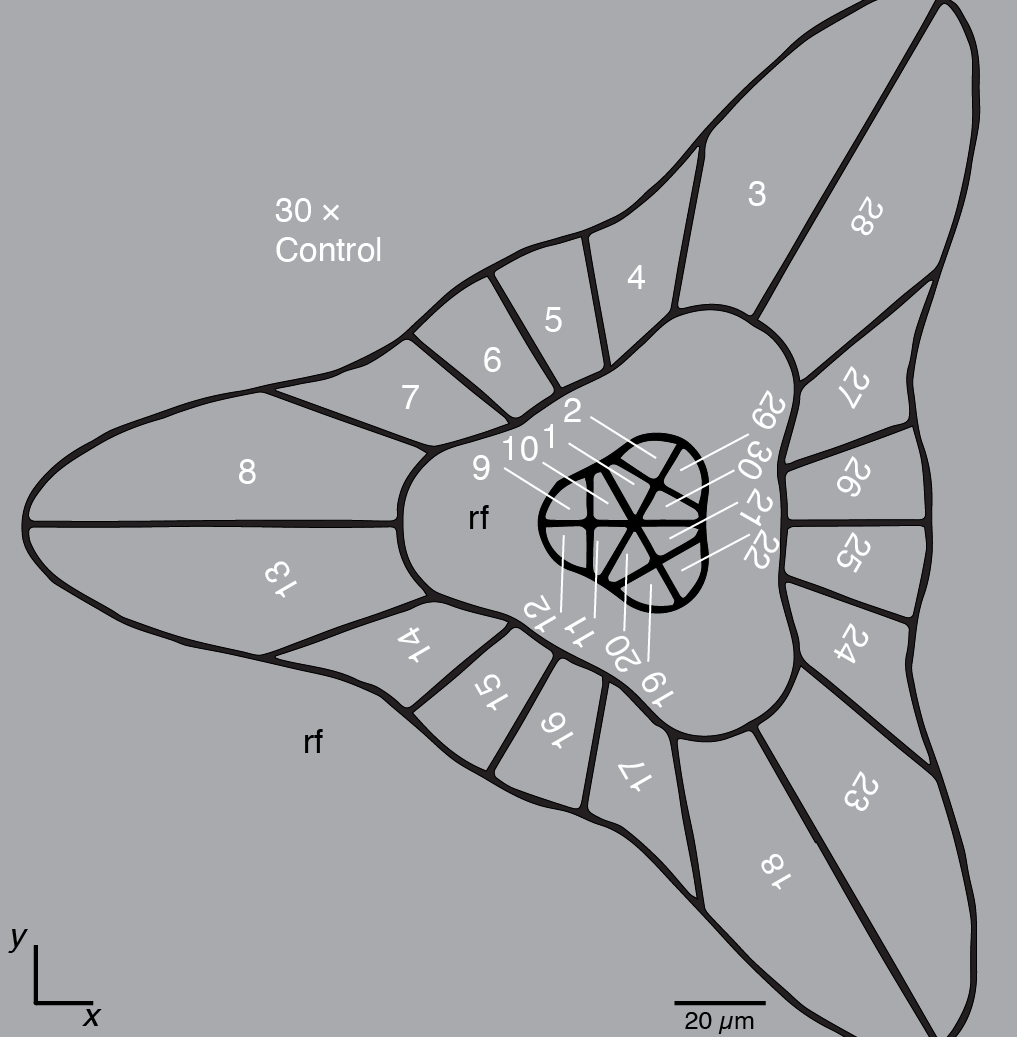}
\caption{\label{fig5}
 Schematic of the central region of our rf surface electrode trap. We show the two rf electrodes and labels of all 30 control electrodes. 
}
\end{figure}

We show a schematic of our trap architecture with the 30 control electrodes in Fig.\,\ref{fig5}; the metal sheet is not shown. 
Using the AWG, we apply voltages corresponding to individual $\Phi_\text{c}$ to all electrodes (31 in total), and list values in Table\,\ref{tab:table2}.
To establish shuttling of single ions from T$_\text{H}$ to all sites, we perform dedicated calibration sequences with single $^{24}$Mg ions and optimize $\Phi_{\text{c}}$ for fixed rf drive parameters.
The procedure comprises two steps:
Step\,1, for shuttling an ion, we linearly ramp between two control potential configurations used for ion storage at two different sites (T$_{\text{H}}$ and the site of choice).
Step\,2, by comparing the collected photon count histograms, we fine-tune individual sets of control potentials to optimize/reduce the overall slew rate needed. Thereby, minimizing the fluorescence loss due to relocation.
We ensure the success of operations via camera images. 
In Table\,I of the main text, we list the maximum slew rates required for the playbacks.
For a single relocation, we play back waveforms between dedicated fine-tuned control potential configurations with minimized slew rates in a time period of 0.1\,ms.
\begin{table}
\caption{\label{tab:table2}Voltages (V) applied to electrodes for dedicated $\Phi_\text{C}$. 
Our AWG has a discretization limit at $3\times10^{-4}$ V (for each channel), and provides voltages in the range of [-10,10]\,V.}
\begin{ruledtabular}
\begin{tabular}{ccccccc}
Electrodes &$\Phi_\text{0}$ &$\Phi_{1}$ &$\Phi_{2}$ &$\Phi_\text{H}$ &$\Phi_\text{HL}$ &$\Phi_\text{Pyramid}$\\
\hline
el1 &0.0959 &0.0959 &0.0959 &0.0959 &-0.0038 &0.0959\\
el2 &-0.1676 &-0.1676 &-0.1676 &-0.1676 &0.1484 &-0.1676\\ 
el3 &-0.2760 &-0.2760 &-0.2760 &-0.2760 &-0.1484 &-0.2760\\ 
el4 &0.2328 &0.2328 &0.2328 &0.2328 &0.1484 &0.2328\\
el5 &-0.1246 &-0.1246 &-0.1246 &-0.0246 &0.0870 &-0.1246\\ 
el6 &0.1520 &0.1520 &0.1520 &0.1520 &-0.1484 &0.1520\\ 
el7 &-0.5683 &-0.5683 &-0.5683 &-0.5683 &-0.1484 &-0.5683\\ 
el8 &0.3005 &0.3005 &0.3005 &0.3005 &-0.0094 &0.3005\\ 
el9 &-0.4945 &-0.4945 &-0.4945 &-0.4945 &-0.1484 &-0.4945\\ 
el10 &-0.4502 &-0.4502 &-0.4502 &-0.4502 &-0.0641 &-0.4502\\ 
el11 &0.7980 &0.7980 &0.7980 &0.7980 &0.1484 &0.7980\\ 
el12 &0.1130 &0.1130 &0.1130 &0.1130 &0.1484 &0.1130\\ 
el13 &0.0327 &0.0327 &0.0327  &0.0327 &0.1484 &0.0327\\ 
el14 &-0.1239 &-0.3239 &-0.1239 &-0.1239 &0.1484 &-0.3239\\ 
el15 &0.0007 &0.0007 &0.0007 &0.0007 &-0.0387 &0.0007\\ 
el16 &0.6000 &0.6000 &0.6000 &0.6000 &-0.1484 &0.6000\\ 
el17 &0.0064  &0.0064 &0.0064 &0.0064 &-0.1484 &0.0064\\ 
el18 &-0.6718 &-0.6718 &-0.6718 &-0.6718 &-0.1484 &-0.6718\\ 
el19 &-0.3833 &-0.3833 &-0.3833 &-0.3833 &-0.1484 &-0.3833\\ 
el20 &-0.0137 &-0.0137 &-0.0137 &-0.0137 &-0.1484 &-0.0137\\ 
el21 &0.1892 &-0.1108  &-0.1108 &-0.1108 &-0.1484 &0.1892\\ 
el22 &-0.1364 &-0.1364 &-0.1364 &-0.1364 &0.0671 &-0.1364\\ 
el23 &0.5604 &0.5604  &0.2604 &0.5604 &0.1484 &0.2604\\ 
el24 &-0.1949 &-0.1949 &-0.1949 &-0.1949 &-0.1111 &-0.1949\\ 
el25 &0.4670 &0.4670 &0.4670 &0.4670 &-0.1484 &0.4670\\ 
el26 &-0.4358 &-0.4358 &-0.4358 &-0.4358 &0.0103 &-0.4358\\ 
el27 &0.1844 &0.1844 &0.1844 &0.1844 &0.1484 &0.1844\\ 
el28 &0.3901 &0.3901 &0.3901 &0.3901 &0.1484 &0.3901\\ 
el29 &-0.6760 &-0.6760 &-0.6760 &-0.6760 &-0.1484 &-0.6760\\ 
el30 &-0.6392 &-0.6392 &-0.6392 &-0.6392 &-0.1484 &-0.6392\\
Metal sheet &-3.5000 &-3.5000 &-3.5000 &-3.5000 &-2.7000 &-3.5000\\
\end{tabular}
\end{ruledtabular}
\end{table}
\end{document}